\documentclass[twocolumn,showpacs,preprintnumbers,amsmath,amssymb]{revtex4}

\usepackage{graphicx}
\usepackage{dcolumn}
\usepackage{bm}

\begin{document}

\title{Magnetic avalanches of minor fast-relaxing species of Mn$_{12}$ acetate}

\author{S. McHugh}
\author{R. Jaafar}
\author{M. P.  Sarachik}
 \email{sarachik@sci.ccny.cuny.edu}
\affiliation{
Department of Physics\\ City College of New York, CUNY\\
New York, New York 10031, USA}

\author{Y. Myasoedov}
\author{A. Finkler}
\author{E. Zeldov}
\affiliation{
Department of Condensed Matter Physics\\
The Weizmann Institute of Science\\
Rehovot 76100, Israel}

\author{R. Bagai}
\author{G. Christou}
\affiliation{
Department of Chemistry\\
University of Florida\\
 Gainesville, Florida 32611, USA}

\begin{abstract}
Using micron-sized thermometers and Hall bars, we report time resolved studies of the local temperature and local magnetization for two types of magnetic avalanches (abrupt spin reversals) in the molecular magnet Mn$_{12}$-acetate, corresponding to avalanches of the main slow-relaxing crystalline form and avalanches of the fast-relaxing minor species that exists in all as-grown crystals of this material.  An experimental protocol is used that allows the study of each type of avalanche without triggering avalanches in the other, and of both types of avalanches simultaneously.  In samples prepared magnetically to enable both types of avalanches, minor species avalanches are found to act as a catalyst for the major species avalanches.
\end{abstract}

                     
\maketitle

\section{\label{sec:level1}Introduction}
In slow chemical combustion, or deflagration, an exothermic chemical reaction takes place along a front that travels with subsonic speed, changing the initial unreacted material ahead of the front into reaction products, or ash, behind.  The process of deflagration is realized in many other contexts and in phenomena that span many orders of magnitude in energy.  As exotic high-energy examples, deflagration describes the burning of carbon in supernovae \cite{Supernovae} and is a proposed mechanism for the burning of a neutron star into strange matter \cite{Strange}.  It has even been used to describe the production of gravitational waves resulting from a vacuum phase transition in the early universe \cite{Gravity}.   As perhaps the lowest energy realization, ``magnetic deflagration" has recently been invoked to account for the magnetic avalanches, or abrupt magnetization reversals, that often occur in crystals of molecular magnets.  Here, Zeeman energy is released \cite{Suzuki, Garanin} when metastable spins opposing an applied magnetic field are flipped in the direction of the field producing a ``flame" front that travels at subsonic speed.  This process is tunable, reversible, and arguably the cleanest manifestation of deflagration.  Most strikingly, since molecular magnets are known to exhibit enhanced relaxation rates due to quantum tunneling \cite{Friedman}, there is the prospect of investigating ``quantum magnetic deflagration" \cite{AlbertoI, AlbertoII, McHughI, McHughII}.

Mn$_{12}$-ac is the prototypical molecular magnet composed of twelve Mn atoms coupled by superexchange to form superparamagnetic clusters of spin S =10 at low temperatures. Arranged in a body-centered tetragonal lattice, the molecules are well separated from each other so that the magnetic interaction between them is negligible and, to a good approximation, the magnetic clusters respond independently to an external magnetic field.  There is a strong magnetic anisotropy along the symmetry (c axis) of the crystal that creates a barrier against spin reversal of approximately 60 K, a barrier that can be reduced by applying a magnetic field along the easy axis of magnetization \cite{Lis, Sessoli}.  Each molecule in the crystal can be modeled with the effective Hamiltonian
\begin{eqnarray}
\mathcal{H} = -DS_z^2 - AS_z^4 -g\mu_BS_zB_z + \mathcal{H}_\perp,
\label{Hamiltonian}
\end{eqnarray}
where $D = 0.548$ K, $A = 1.17 \times 10^{-3}$ K, $g = 1.94$, and $\mathcal{H}_\perp$ is a small symmetry-breaking term that gives rise to tunneling \cite{g Factor, Hamiltonian};  $S_z$ is the component of the spin lying along the c-axis of the crystal and the magnetic field $B_z  = \mu_0(H_z + M_z)$, where $H_z$ is the applied magnetic field and $M_z$ is the magnetization.

The anisotropy barrier allows the spins to be prepared in a metastable state anti-parallel to an external magnetic field.  When a molecule's spin reverses, making a transition from the metastable well to the ground state of the stable well, the Zeeman energy is released to phonons (heat) which diffuse to neighboring molecules and thermally stimulate further reversal.  Given the appropriate conditions, a thermal runaway can occur resulting in the abrupt complete reversal of the crystal's magnetization.  Paulsen and Park \cite{Paulsen} were the first to report these magnetic avalanches and to propose a generic thermal runaway as an explanation.  Since then, the thermal nature has been confirmed in a number of studies (see, for example, Refs.  \cite{Fominaya, Tejada, Alberto Thermal, Webster}).   In 2005, Suzuki et al.  \cite{Suzuki} discovered that the spin reversal proceeds in the form of a deflagration front.  A comprehensive theory of magnetic deflagration has now been developed by Garanin and Chudnovsky \cite{Garanin}.  

It is well-known that all single crystals of Mn$_{12}$-ac contain two ``species" of molecules.  The primary or ``major" species described above comprises roughly 95\% of the crystal and relaxes toward equilibrium slowly due to the high anisotropy barrier.  The remaining magnetic clusters, a secondary or ``minor" species, are  low-symmetry, fast-relaxing molecules \cite{minorRef, WernsdorferI}.  Although not as thoroughly characterized as the major species, the minor species molecules can also be modeled with a similar effective spin Hamiltonian
\begin{eqnarray}
\mathcal{H} \approx -dS_z^2 -g\mu_BS_zB_z + \mathcal{H}_\perp,
\label{Minor Hamiltonian}
\end{eqnarray}
with a lower anisotropy barrier $d = 0.49$ K, while the g-value and spin remain the same: $g=1.94$, and $S = 10$ \cite{minorRef, WernsdorferI, WernsdorferII, McHugh Dipole}.

In this paper we report the results of a detailed study of the magnetization dynamics and conditions for ignition of avalanches of both species of Mn$_{12}$-ac.  By employing an experimental protocol to magnetically stabilize one or the other species \cite{WernsdorferII, McHugh Dipole}, we are able to study avalanches of either the major species exclusively or the minor species exclusively: so-called ``major avalanches" and ``minor avalanches", respectively.  We also studied avalanches where both species are allowed to relax during an avalanche, the so-called ``combined avlanches".  We find that the presence of the minor species acts to simultaneously lower the ignition threshold and increase the avalanche speed.  This is analogous to  a chemical catalyst acting to increase a chemical reaction rate, thereby deepening the connection between chemical and magnetic deflagration.  Finally, we apply the theory of magnetic deflagration to the minor and major avalanches for a quantitative analysis.  

\begin{figure}[htbp]
\begin{center}
\includegraphics[height=1.35in]{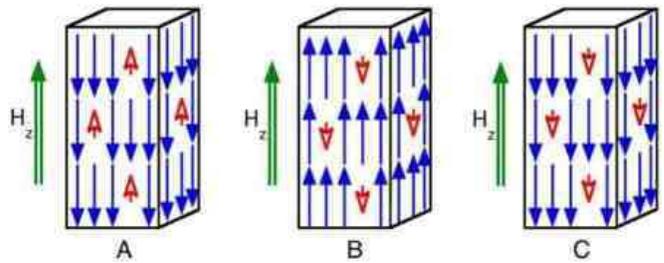}
\caption{(Color online) Schematic crystal for the three different classes of avalanches: A) ``Major avalanches"; only the major species reverses.  B) ``Minor avalanches"; only the minor species reverses.  C) ``Combined avalanches"; both species reverse during avalanche.  The long arrows represent the major species molecules and the short open arrows represent the minor species.}
\label{fig1}
\end{center}
\end{figure}

\section{\label{sec:level1}Experiment and results}

\begin{figure}[htbp]
\begin{center}
  \includegraphics[width=3in, height=5in]{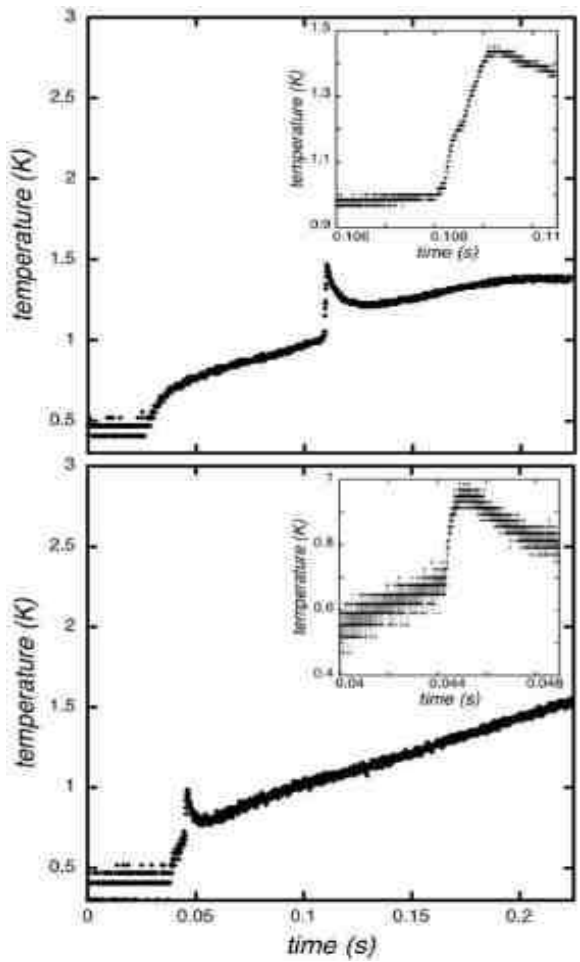}
\caption{Temperature recorded by thermometer in contact with crystal during the triggering of an avalanche.  (a) Major avalanche triggered at $0.83$T. The heater is turned on at $0.03$s and the temperature begins to increase slowly.  The abrupt rise in temperature at $0.11$s is due to heat released by the avalanche. The inset shows data taken at the ignition temperature with higher resolution for the same avalanche.  (b) Minor avalanche triggered at $0.83$T \cite{foot}.}
\label{fig2}
\end{center}
\end{figure}

\subsection{Magnetic Preparation}
The relaxation rates of both species are well described with an Arrhenius law, 
\begin{eqnarray}
\Gamma = \Gamma_0 \mbox{exp}\left(-U(H)/T\right),
\label{arrhenius}
\end{eqnarray}  
where $\Gamma_0 = 3.6\times 10^7$ s$^{-1}$ for the major species \cite{Gomes} and $\Gamma_0 = 4.5\times 10^9$ s$^{-1}$ for the minor species \cite{Soler}.
Although both species have the same spin value, $S=10$, the anisotropy barrier of the minor species is lower.  As a consequence, a smaller external field ($1.5$ T) is required to reverse the minor species at low temperatures ($300$ mK).  As described below (and in Refs. \cite{WernsdorferII, McHugh Dipole}), this allows a sample to be prepared in which the major and minor species have anti-parallel spin alignment.  In turn, this enables us to investigate minor and major avalanches separately.

To accomplish this, the sample is first completely magnetized in the ``down" direction with a large ($-5$ T) external magnetic field; the field is then swept at $+5$ mT/s in the opposite ``up" direction  to a value that is large enough to reverse the minor species upward but small enough to leave the major species unchanged.  At $300$ mK, $+2.0$T is sufficient to reverse the minor species while leaving the major species metastable.  Bringing the magnetic field back to zero yields a crystal with the major and minor species fully magnetized in opposite directions.  

We take advantage of this magnetic partitioning to study avalanches consisting of exclusively one species or the other species.  We studied three classes of avalanches, depicted schematically in Fig. \ref{fig1}.  The long filled arrows represent the major species and the short open arrows represent the minor species.  The external field $H_z$ is shown as the long double arrow outside the crystal.  The major avalanche preparation is shown in A, where only the major species is anti-parallel to the applied field and therefore, metastable.  Avalanches involving the minor species alone (minor avalanches) are shown in B.  

We also studied combined avalanches where both species relax, as shown in Fig. \ref{fig1} C.  Combined avalanches, those in which both species relax, are the simplest to prepare.  After magnetizing both species at $-5.0$ T, one simply triggers an avalanche at a positive field less than $+2.0$ T, the minor species' coercive field.

\begin{figure}[htbp]
\begin{center}
  \includegraphics[width=3in, height=5in]{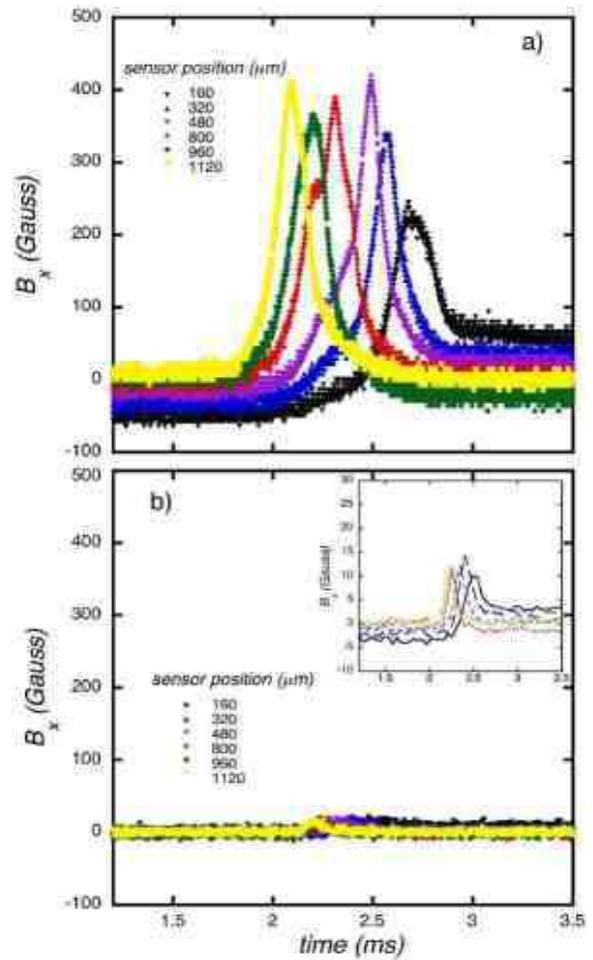}
\caption{(Color online) Signals recorded by six Hall sensors in contact with the crystal during an avalanche. (a) Major species  triggered at $0.9$T.  The velocity of the avalanche is determined by recording the arrival time of the peak at each of the six sensors. (b) Minor species avalanche triggered at $0.9$T.  The inset shows the output of each sensor with the high frequency noise averaged out.}
\label{fig3}
\end{center}
\end{figure}

\begin{figure}[htbp]
\begin{center}
  \includegraphics[width=3in, height=5in]{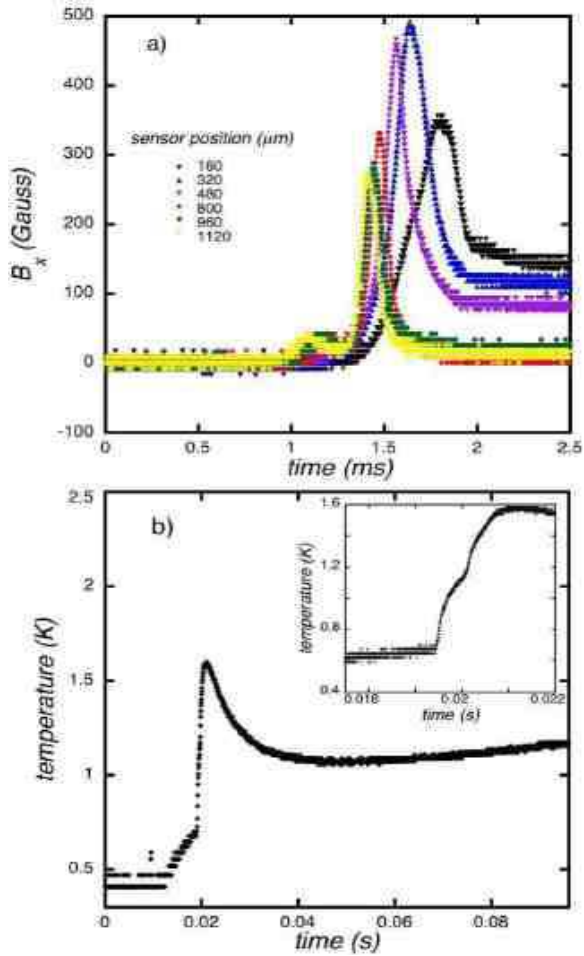}
\caption{(Color online) Avalanche dynamics for a combined avalanche triggered at $0.83$ T.  (a) Hall sensor responses.  (b) Temperature profiles.  The inset displays the data at ignition with greater resolution.}
\label{fig4}
\end{center}
\end{figure}

\begin{figure}[htbp]
\begin{center}
  \includegraphics[height=5in]{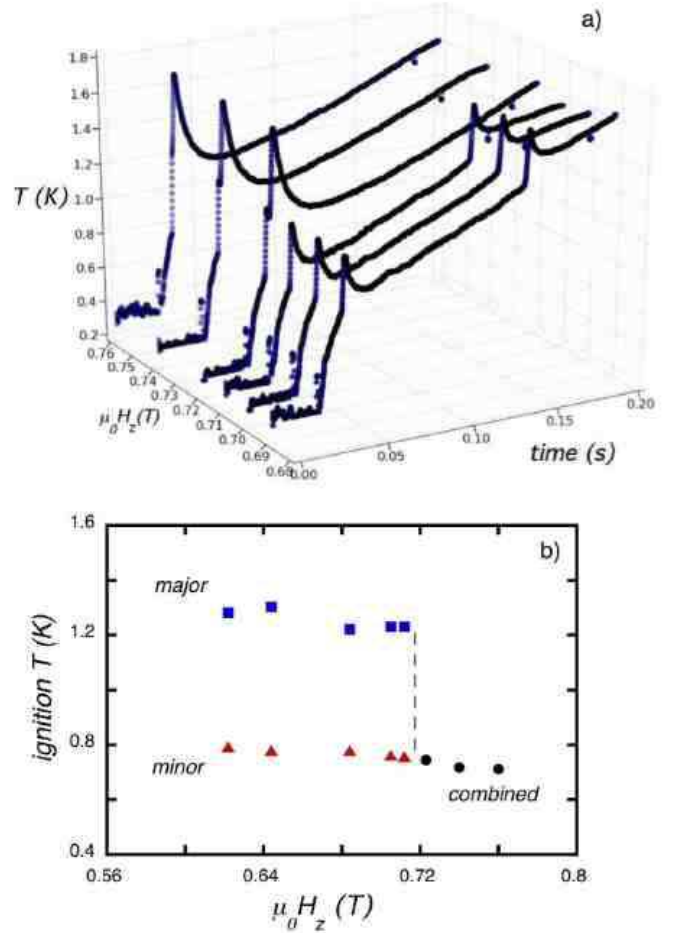}
\caption{(a)  Temperature profiles for combined avalanches triggered at low fields. (b) Ignition temperature as a function of magnetic field.  The minor and major species avalanches are triggered separately below a sample-dependent magnetic field, while at higher fields ignition of the minor species triggers the ignition of the major species.  The transition from one regime to the other is discontinuous.}
\label{fig5}
\end{center}
\end{figure}

\begin{figure}[htbp]
\begin{center}
  \includegraphics[width=3.25in, height=5.5in]{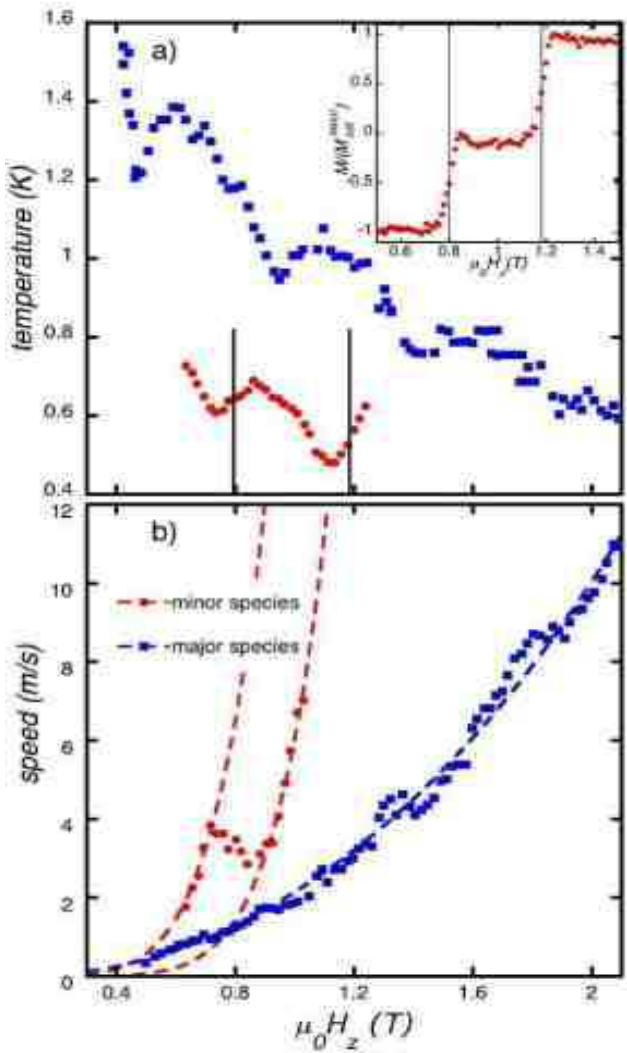}
\caption{(Color online) Comparison of major and minor species avalanches. (a) Ignition temperatures.  The major species avalanches (squares, upper curve) display clear minima in the ignition temperatures at the resonant fields where quantum tunneling occurs.  The inset shows a portion of the hysteresis curve for the minor species, where $M^{minor}_{sat}$ is the full magnetization of the minor species molecules and the vertical lines are drawn at the center of the tunneling resonances.  Ignition minima are observed for the minor species that do not coincide with the minor species resonant fields (the vertical lines drawn in the main part of the figure) but are due instead to loss of magnetization prior to avalanche ignition, as discussed in detail in the text.    (b) Avalanche speeds for the major species (lower curve) and minor species (upper curve).  The dashed lines are approximate fits to the theory of magnetic deflagration of Garanin and Chudvnosky\cite{Garanin}.  The fit for the major species omits the data at resonant magnetic fields, where the avalanche speeds are enhanced.  For the minor species avalanches the left-hand curve assumes  $\Delta M/2M_{sat} = 0.05$ and the right-hand curve (above $0.85$ T) is for $\Delta M/2M_{sat} = 0.025$, as explained in the text.}
\label{fig6}
\end{center}
\end{figure}

\begin{figure}[htbp]
\begin{center}
  \includegraphics[width=3in, height=6.5in]{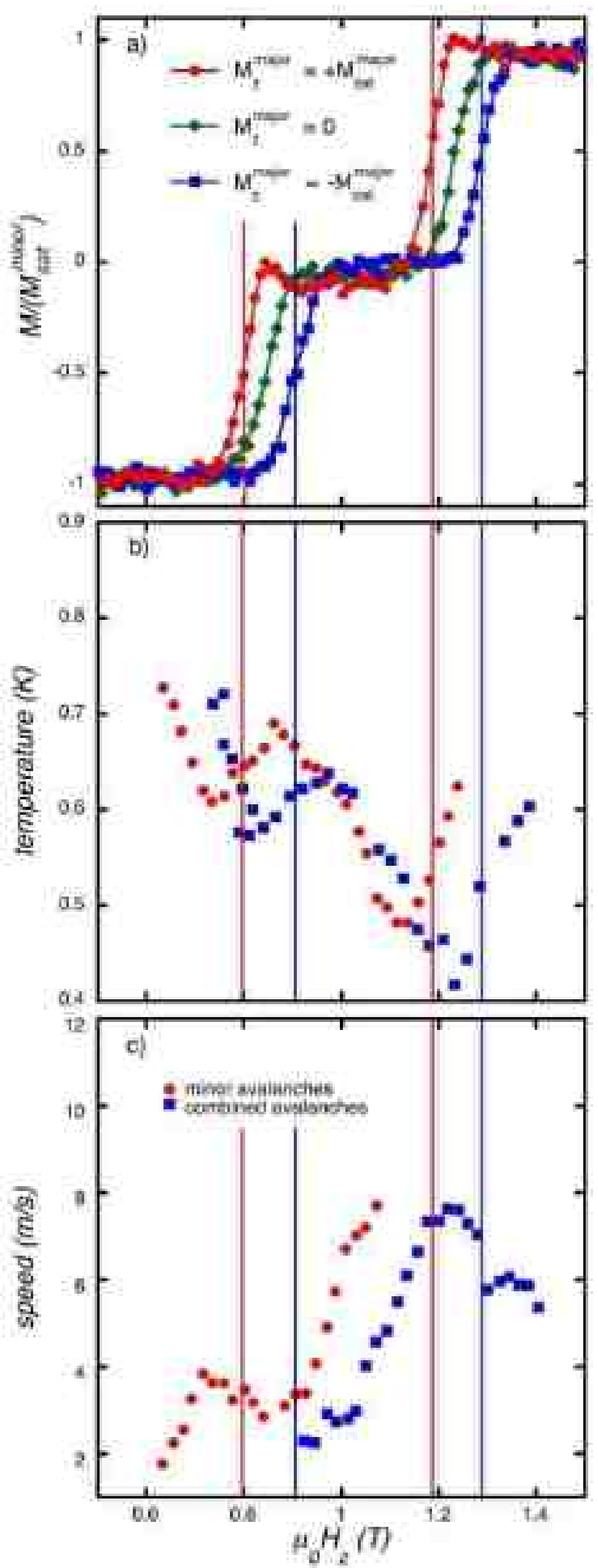}
\caption{(Color online) (a) A portion of the normalized hysteresis curve is shown for minor species, with the major species prepared in three different magnetizations: $\pm M^{major}_{sat}$ and $0$, where $M^{major}_{sat}$ is defined as the magnetization due to the major species when completely aligned.  The shift of the resonance fields by $\approx \pm 0.05$ T from the zero magnetization case is due to the dipole field of the major species.  The structure of the (b) ignition temperatures and (c) speeds of the minor and combined avalanches also reflects this minor species resonance field shift. }
\label{fig7}
\end{center}
\end{figure}

\subsection{Procedure and results}
All measurements reported here were performed on single crystals of Mn$_{12}$-ac with typical dimensions $1.5 \times 0.3 \times 0.3$ mm$^3$ immersed in liquid $^3$He at approximately $300$ mK.  For temperature measurements, germanium thin film resistance thermometers of dimensions $40 \times 100$ $ \mu$m$^2$ were deposited by e-gun evaporation through shim masks onto heated GaAs substrates in vacuum.  For the magnetic measurements, six micron-sized Hall sensors, fabricated using two-dimensional-electron-gas in GaAs/AlGaAs heterostructures, were used to record the local magnetization as a function of time.  The crystal was mounted on the thermometer (for measurements of the temperature) or on the Hall sensor array (for magnetization measurements) using a thin layer of thermally conductive Apiezon M grease.  A twisted wire of Constantan was used as a heater.  In order to make good thermal contact with the heater, the entire assembly, including thermometer (or Hall sensors), sample, and heater, was encased in Apiezon M grease.  The heater was placed roughly $1$ mm above the crystal.  The minimum heater power was used that still triggered avalanches.

With the sample suitably prepared following one of the magnetic protocols described above, an avalanche was triggered by applying a constant current through the heater.  Figure \ref{fig2} (a) shows the response of the thermometer for an avalanche of the major species.  The heater was turned on at $0.03$ s, and the subsequent slow rise in temperature from $t \approx 0.03$ to $t \approx 0.11$ s reflects the gradual heating of the entire sample in response to the power provided by the heater (which remains on).  The sharp rise in temperature at $t \approx 0.11$ seconds signals the sudden release of heat associated with the ignition of an avalanche \cite{foot}.  The temperature at which this occurs is denoted as the ignition temperature.  Figure \ref{fig2} (b) shows the thermometer response for an avalanche of the minor species.

Measurements were repeated several times at a given field, and were reproducible within a given run.  The ignition temperature was reproducible within a given experimental run, but varied by as much as $0.25$ K from one run to another.  This is most likely due to uncontrolled thermal gradients that were different depending on the thermal connection between the thermometer and the sample.  For example, the thickness of the layer of Apiezon M grease was perforce different for different runs.  It may also be due to sample geometry (shape and size) which allows the heat generated by the avalanche to leave the crystal more or less easily.

In separate experimental runs for similar size Mn$_{12}$-ac crystals, the magnetization dynamics were measured with the Hall sensor array using an excitation current of $66$ $\mu$A.  As in earlier experiments \cite{Suzuki, McHughI}, this enabled us to track the propagation of the narrow interface between regions of the sample with anti-parallel magnetizations corresponding to the magnetic avalanche.  Figure \ref{fig3}  shows the output of the six Hall sensors for an avalanche triggered at $0.83$ T.  The velocity of the deflagration front was deduced from measurements of the arrival time of the peak at each sensor.  Figures \ref{fig3} (a) and \ref{fig3} (b) show typical outputs for a major avalanche and a minor avalanche, repectively.  Because of the significantly smaller signal produced by the minor avalanche, signal averaging was applied to remove high frequency noise.  The result is shown in the inset.  The ratio of the minor peak amplitude to that of the major peak is about $0.04$, in reasonable agreement with the expected ratio of minor species to major species in the sample, $\approx 0.05$. 

Combined avalanches were triggered with both the major and minor species reversing magnetization during the avalanche (see Fig. \ref{fig1} C).  Figure \ref{fig4} shows data for the time-dependence of the temperature and the local magnetization for such an avalanche triggered at $0.83$ T.  At low fields, the two species do not relax simultaneously during an avalanche.  Figure \ref{fig5} shows the separation of the two species as reflected in the temperature profiles.  Below roughly 0.73 T, the minor species relaxes prior to and independently of the major species.  The precise field below which this occurs varies from crystal to crystal.  

Figure \ref{fig6} shows the ignition temperature and the avalanche speed as a function of magnetic field for avalanches of the major and the minor species, each acting alone.  Figures \ref{fig7} (b) and (c) show the ignition temperatures and speeds for the combined avalanches (with data for the minor species included for comparison).  The speeds and ignition temperatures for the combined avalanches in Fig. \ref{fig7} are shown only for magnetic fields sufficiently high to induce simultaneous avalanche of the two species  (as discussed above).

\section{\label{sec:level1}Discussion}

\subsection{Major and minor avalanches}

The overall behavior of avalanches of the major and the minor species, each acting alone, can largely be understood with the help of Eq. \ref{arrhenius}.  

The blocking temperature is the temperature above which the spins can freely reverse on the time scales relevant to the experiment and is closely related to the relaxation rate.  Turning on the heater results in slow heating of the crystal until the temperature approaches the blocking temperature, at which point the spins can reverse and an avalanche may occur.  Since the relaxation rate is much higher for the minor species, we expect that the ignition threshold should be lower.  This is confirmed by Fig. \ref{fig6} (a) showing much lower ignition temperatures for the minor avalanches than the major species avalanches.  

Similar behavior is evident for the avalanche speeds.  Once an avalanche is ignited, the speed of the front must be related to the rate at which the metastable spins at the deflagration front \emph{react}.  Guided by Eq. \ref{arrhenius}, we expect that the lower barrier $U(H)$ of the minor species should result in faster speeds for the minor avalanches.  Again, this is confirmed by Fig. \ref{fig6} (b).  

For both major and minor species avalanches, the ignition temperatures and the avalanche speeds display behavior that is clearly nonmonotonic.  The minima in the major avalanche ignition temperatures at the fields $0.45$ T, $0.9$ T, and $1.35$ T correspond to a reduction of the major species barrier by quantum tunneling.  This decrease in the ignition threshold was discussed at some length for these data in Ref. \cite{McHughI}.

At first glance one may attribute the minima in the minor ignition temperatures to a similar process of quantum tunneling at the resonant fields relevant to the minor species.  However, this is not so.  The ignition temperature minima for the minor species are due instead to the loss of magnetization prior to the ignition of the avalanche.  To make this clear, the inset of Fig. \ref{fig6} (a) shows a quarter of the minor species hysteresis curve taken with the major species magnetized in the positive direction (data taken from Ref. \cite{McHugh Dipole}).  As the field is swept at $+5$ mT/s across the minor species resonance fields at $0.79$ T and $1.20$ T, a portion of the metastable magnetization has relaxed, thus removing some of the \emph{fuel} available for deflagration.  The vertical black lines drawn on Fig. \ref{fig6} indicate the center of the minor species resonance fields.  For fields below $0.74$ T, the entire minor species magnetization reverses during an avalanche.  However,  for those triggered above $0.85$ T and up to $1.15$ T, only about half the magnetization reverses during the avalanche.  Finally, for fields above $1.20$ T, all the minor species magnetization has relaxed, precluding the possibility of igniting an avalanche.  

Similarly, the avalanche speeds for the minor species exhibit behavior that should not be confused with quantum enhancement of the avalanche speed.  The apparent peak in the minor avalanche speeds occurs at a lower field than the minor species tunneling resonance.  As the field is increased and approaches the minor species resonance near $0.79$ T, a portion of the minor species magnetization relaxes prior to the ignition of the avalanche.  This loss of magnetization is equivalent to the loss of fuel available to burn in chemical deflagration, which leads to slower speeds.

Non-monotonic behavior is also observed for the avalanche speeds of the \emph{major} avalanches.  Although small at low fields, there is an apparent increase of speed at $1.35$ T and $1.80$ T, consistent with quantum tunneling of the majority spins.  These major avalanche data were presented and discussed in detail in Ref. \cite{McHughI}.  It is worth noting that a similar quantum enhancement of the avalanche speed should also be present for the minor avalanches; however, it is overwhelmed by the effect of the loss of magnetization prior to ignition.  In order to observe the quantum enhancement of avalanche speeds in either species of avalanche, one must create conditions that eliminate the pre-ignition losses by preparing the magnetization at lower temperatures or sweeping the external field at a faster rate.

The theory of magnetic deflagration of Garanin and Chudnvosky \cite{Garanin} gives an approximate formula for the speed of propagation of a planar deflagration front, 
\begin{eqnarray}
v = \sqrt{\frac{3k_BT_f\kappa \Gamma(U, T_f)}{U(B)}},
\label{EOM}
\end{eqnarray}
where $T_f$ is the temperature reached at the front, $\kappa$ is the thermal diffusivity, $\Gamma$ is the spin relaxation rate (Eq. \ref{arrhenius}), and $U(B)$ is the barrier against spin reversal.  Applying procedures reported in Ref. \cite{McHughII}, we can obtain a fit to the theory for the major species avalanches as follows. 

In order to calculate $T_f$ for a particular avalanche, we require the average energy released per molecule and the heat capacity.  The average energy released per molecule as it relaxes is proportional to the Zeeman energy, 
\begin{eqnarray}
\langle E\rangle = 2 g\mu_B S B_z\left(\frac{\Delta M}{2M_{sat}}\right).  
\end{eqnarray}
The parameter $\Delta M/2M_{sat}$ is introduced to account for avalanches with different fractions of initial metastable spins.  For instance, if all spins (both major and minor) relax during the avalanche, then $\Delta M/2M_{sat} = 1$; if the fully magnetized major only (minor only) species relaxes during an avalanche then $\Delta M/2M_{sat} = 0.95$ $(0.05)$.  Given $\langle E \rangle$, the maximum possible temperature ($T_{max}$) during an avalanche is calculated as the upper limit of the following integral:
\begin{eqnarray}
\langle E\rangle = \int_0^{T_{max}} dT C(T, B_z),
\label{heatCap}
\end{eqnarray}
where $C(T, B_z)$ is the specific heat capacity per molecule \cite{Gomes, McHughII}.  For the major species avalanches, values of $T_{max}$ are obtained that range from $8$ to $14$ K.  Since the density of minor species is small, $T_{max}$ for the minor species avalanches ranges between $3$ and $4$ K.  We assume the heat loss through the surfaces of the crystal are negligible, therefore $T_f = T_{max}$.  

The barrier for the major species molecules, $U_{major}$, is calculated from the spin Hamiltonian, Eq. \ref{Hamiltonian}.  The magnetization, $M_z$, produces an internal field of $51.5$ mT that has a clear effect on the avalanche speeds (as evidenced by Fig. \ref{fig7}).  We take into account its effect on $U_{major}$ by setting $B_z =( \mu_0H_z - 51.5)$ mT, where $B_z$ is the externally applied field.

Omitting the data at resonant magnetic fields (where the avalanche speeds are enhanced by quantum tunneling) these calculated values for the major species avalanches are used to obtain the fit to Eq. \ref{EOM} shown in Fig. \ref{fig6} (b).  The thermal diffusivity was allowed to vary with temperature as $\kappa = aT_f^\alpha$; based on results obtained by applying this analysis to an extensive data set (see \cite{McHughII}), we set the exponent $\alpha = 3$ and obtained a ``best" fit for $\kappa = 1.0\times10^{-7} \times T_f^{3}$ m$^2$/s.  

A similar procedure used to fit the available data for the minor species avalanches yields a thermal diffusivity that has a temperature dependence $\propto T_f^x$ with $x \approx 7$.  As shown by the dashed lines in Figure \ref{fig6} (b), two separate fits are applied: at low fields, all the minor species molecules are available and relax during the avalanche, and we estimate that $\Delta M/2M_{sat} = 0.05$, while at higher fields approximately half the minor spins have relaxed prior to ignition and $\Delta M/2M_{sat} = 0.025$.

There is a large error associated with applying this analysis to the minor species, as the minor species spin density is known only approximately and there are fewer data points than for the major species.  It is nevertheless clear that the temperature dependence one deduces by fitting to the theory is sharply positive, with a temperature exponent that is substantially greater than $T_f^3$.  This is quite unexpected, as thermal diffusivity generally decreases with increasing temperature \cite{LowTemperaturePhysics, Poiseuille}.  The fact that, for both major and minor species avalanches, fits to the theory of magnetic deflagration of Garanin and Chudnovsky \cite{Garanin} produce thermal diffusivities that $increase$ with temperature suggests that assumptions we used in the analysis may be unwarranted, or that the theory is incomplete in its present form.

\subsection{Combined avalanches}
We now examine the data obtained for the combined avalanches, where both major and minor species relax during the avalanche.  In Fig. \ref{fig4}, the magnetic data for an avalanche triggered at $0.83$T show that a small amount of relaxation precedes the large peaks due to the major species avalanche.  This precursor is the reversal of the spin magnetization of the minor species just before that of the major species.   Figure \ref{fig5} shows the separation of the two species as reflected in the temperature profiles.  For all samples measured, the two species do not relax simultaneously during an avalanche below a field on the order of $0.75$T which varies from crystal to crystal.  At low fields the minor species relaxes prior to and independently of the major species, while above this field the major species and minor species ignite together and propagate as a single front; the transition between these two regimes appears to be discontinuous.  It is analogous to grass and trees that can sustain separate burn fronts that abruptly merge into a single front when the grass becomes sufficiently hot to ignite the trees.  This  interesting behavior warrants further investigation \cite{separatespeed}.  

The presence of the minor species in the combined avalanches significantly lowers the ignition threshold as well as increasing the avalanche speed.  In fact, as shown in Figs. \ref{fig7} (b) and (c), the ignition temperatures and speeds for the combined avalanches are nearly equal to those of the minor avalanches.  This suggests that the presence of the metastable minor species acts as a catalyst for the major species in analogy with the behavior of a catalyst in a chemical reaction.  

A close look at the ignition temperatures of the minor avalanches and the combined avalanches in Fig. \ref{fig7} (b), reveals similar minima.  This is due to the loss of minor species magnetization prior to the ignition of the avalanche.  However, the minima occur at different fields.  In the following section, we make the analogy with a catalyst more complete by explaining the shifts in the resonant fields between the minor and the combined avalanches.

\subsection{Effects of major species dipolar field}

The shift in the minor species' resonant fields can be explained by carefully taking into account the dipolar field due to the major species.  Figure \ref{fig7} (a), taken from Ref. \cite{McHugh Dipole}, shows a quarter of the hysteresis curve for the minor species for three different magnetizations of the \emph{major} species.  The squares are data taken with major species completely magnetized in the negative direction, $M_z^{major} = -M_{sat}$.  The circles are data taken with the major species magnetized in the positive direction, $M_z^{major} = + M_{sat}$.  Finally, the diamonds denote data taken with the major species randomly oriented in the positive and negative directions, thus giving zero net magnetization of the major species, $M_z^{major} = 0$.  All curves were taken at a temperature of $0.3$ K and an external field sweep rate of $5$ mT/s.  

The locations of the tunneling resonances for the minor species are determined by the total field, $B_z = \mu_0(H_z + M_z^{major})$ \cite{footII}.  The shift of the minor species resonance field relative to the middle curve is a direct measure of the dipolar field due to the fully magnetized major species, $M_z^{major} = \pm M_{sat}$, where $\pm \mu_0M_{sat} = \pm 51.5$ mT.  

The shift due to $M_z^{major}$ is also reflected in the ignition temperatures and avalanche speeds shown in Figs. \ref{fig7} (b) and (c).  Figure \ref{fig7} (b) shows the ignition temperatures of the combined (squares) and minor (circles) avalanches.  Similarly, Fig. \ref{fig7} (c) is a comparison between the avalanche speeds.  In both (b) and (c), the minima and maxima are displaced by $\approx 100$ mT from one another.  Prior to the ignition of minor avalanches, the major species is anti-parallel contributing an additional $\mu_0M_z^{major}\approx +50$ mT to the effective magnetic field, $B_z$, applied to the minor species.  The minor species resonant behavior shifts $-50$ mT from the applied field accordingly.  Prior to the ignition of combined avalanches, the major and minor species are parallel, with the major species contributing $\mu_0M_z^{major}\approx - 50$ mT to $B_z$.  Since the characteristics of the minor species controls the behavior of the combined avalanches, the effective field on the minor species is shifted $+50$ mT.  Again, both circumstances are consistent with the shift of the minor species resonances shown in Fig. \ref{fig7} (a) \cite{McHugh Dipole}.

\section{\label{sec:level1}Summary and Conclusions}

We have observed two species of avalanches in a Mn$_{12}$-ac crystal corresponding to the major and minor species within the sample characterized by different anisotropy barriers and relaxation rates.  A protocol is described that enables the study of avalanches involving only the major species and only the minor species as well as avalanches of both species together.  Although it constitutes only 5\% to 7\% of the sample, the fast-relaxing minor species can sustain an avalanche independently, in the absence of participation of the major species.  The speed of the major species avalanche front displays maxima and the ignition temperature displays minima at the magnetic fields that allow quantum tunneling across the anisotropy barrier.   The ignition temperatures and speeds of the minor species avalanches also display non-monotonic behavior as a function of magnetic field.  However, the nonmonotonic behavior in this case is {\em not} associated with quantum tunneling resonances.  Rather, it is a result of magnetization loss due to quantum tunneling during the magnetic preparation of the sample {\em prior} to triggering of the avalanche.  When both the major and the minor species are allowed to participate in the avalanche, it is found that the fast-relaxing minor species behaves as a catalyst for the deflagration of the major species.

We thank Eugene Chudnovsky and Dmitry Garanin of Lehman College for many illuminating discussions and Yosi Yeshurun of Bar Ilan Universtiy for valuable comments and suggestions.  We are grateful to Hadas Shtrikman of the Weizmann Institute of Science for providing the wafers from which the Hall sensors were made.  This work was supported at City College by NSF grant DMR-00451605.  E. Z. acknowledges the support of the Israel Ministry of Science, Culture and Sports.  Support for G. C. was provided by NSF grant CHE-0414555.


\begin{thebibliography}{5}

\bibitem{Supernovae}
{Ken'ichi Nomoto, Daiichiro Sugimoto and Sadayuki Neo},  Astrophysics and Space Science {\bf 39}, 2, L37 (1976).

\bibitem{Strange}
{G. Lugones, O. G. Benvenuto, and H. Vucetich}, Phys. Rev. D {\bf 50}, 6100 - 6109 (1994)

\bibitem{Gravity}
{Marc Kamionkowski, Arthur Kosowsky, and Michael S. Turner}, Phys. Rev. D {\bf49}, 2837 - 2851 (1994).

\bibitem{Suzuki}
{Yoko Suzuki, M. P. Sarachik, E. M. Chudnovsky, S. McHugh, R. Gonzalez-Rubio, Nurit Avraham, Y. Myasoedov, E. Zeldov, H. Shtrikman, N.E. Chakov, and G. Christou }, Phys. Rev. Lett. {\bf 95}, 147201 (2005).

\bibitem{Garanin}
{D.A. Garanin and E.M. Chudnovsky},Phys. Rev. B {\bf76}, 054410 (2007).

\bibitem{Friedman}
{J. R. Friedman, M. P. Sarachik, J. Tejada, and R. Ziolo}, Phys. Rev. Lett. {\bf 76},  3830  (1996).

\bibitem{AlbertoI} 
{A. Hernandez-Minguez, J. M. Hernandez, F. Macia, A. Garcia-Santiago, J. Tejada, and P. V. Santos}, Phys. Rev. Lett. {\bf 95}, 217205 (2005).

\bibitem{McHughI}
{S. McHugh, R. Jaafar, M. P. Sarachik, Y. Myasoedov, A. Finkler, H. Shtrikman, E. Zeldov, R. Bagai, and G. Christou} , Phys. Rev. B {\bf76}, 172410 (2007).

\bibitem{AlbertoII}
{A. Hernandez-Minguez, F. Macia, J. M. Hernandez, J. Tejada, and P. V. Santos}, J. Magn. and Magn. Matt., 1457-1463, 320 (2008).

\bibitem{McHughII}
{S. McHugh, Bo Wen, Xiang Ma, M. P. Sarachik, Y. Myasoedov, H. Shtrikman, E. Zeldov, R. Bagai, and G. Christou}, http://arxiv.org/abs/0902.0531.

\bibitem{Lis}
{T. Lis}, Acta Cryst. {\bf B 36}, 2042 (1980)

\bibitem{Sessoli}
{R. Sessoli, D. Gatteschi, A. Caneschi, and M. A. Novak}, Nature
(London) {\bf 365}, 141  (1993).

\bibitem{g Factor}
{R. Sessoli,  H.-L.Tsai, A. R. Schake, S. Wang, J. B. Vincent, K. Folting, D. Gatteschi, G. Christou, and D. N. Hendrickson}, J. Am. Chem. Soc. {\bf 115}, 1804 (1993).

\bibitem{Hamiltonian}
{S. Hill, J. A. A. J. Perenboom, N. S. Dalal, T. Hathaway, T. Stalcup, and J. S. Brooks}, Phys. Rev. Lett. {\bf 80}, 2453 (1998), {I. Mirebeau et al.}, Phys. Rev. Lett. {\bf83}, 628 (1999).

\bibitem{Paulsen}
{C. Paulsen and J.G. Park}, in {\it Quantum Tunneling of Magnetization-QTM'94}, edited by L. Gunther and B. Barbara (Kluwer, Dordrecht, The Netherlands, 1995), pp. 189-207. 

\bibitem{Fominaya}
{F. Fominaya, J. Villain, P. Gandit, J. Chaussy, and A. Caneschi}, Phys. Rev. Lett. {\bf 79}, 1126 (1997).

\bibitem{Tejada}
{J. Tejada, E.M. Chudnovsky, J.M. Hernandez, and R. Amigo}, Appl. Phys. Lett. {\bf 84}, 2373 (2004).

\bibitem{Alberto Thermal}
{A. Hernandez-Minguez, A. Jordi, R. Amigo, A. Garcia-Santiago, J.M. Hernandez, and J. Tejada}, Europhys. Lett. {\bf 69}, 270 (2005).

\bibitem{Webster}
{C.H. Webster, O. Kazakova, J.C. Gallop, P.W. Josephs-Franks, A. Hernandez-Minguez, A.Ya. Tzalenchuk}, Phys. Rev. B {\bf 76}, 012403 (2007).

\bibitem{minorRef}
{A. Caneschi et al. }, J. Magn. and Magn. Matt. 177-181, 1330 (1998); {Z. Sun, D. Ruiz, N. R. Dilley, M. Soler, J. Ribas, K. Folting, M. B. Maple, G. Christou and D. N. Hendrickson}, Chem. Commun., (Cambridge), 19, 1973 (1999); 

\bibitem{WernsdorferI}
{Wernsdorfer, R. Sessoli, D. Gatteschi}, Europhys. Lett. 47, 254 (1999). 

\bibitem{WernsdorferII}
{W. Wernsdorfer, N.E. Chakov, G. Christou}, http://arxiv.org/abs/cond-mat/0405014.

\bibitem{McHugh Dipole}
{S. McHugh, R. Jaafar, M. P. Sarachik, Y. Myasoedov, H. Shtrikman, E. Zeldov, R. Bagai, and G. Christou}, Phys. Rev. B {\bf 79}, 052404 (2009).

\bibitem{Gomes}
{A. M. Gomes, M. A. Novak, R. Sessoli, A. Caneschi, and D. Gatteschi}, Phys. Rev. B {\bf57}, 5021(1998).

\bibitem{Soler}
{M. Soler, W. Wernsdorfer, Z. Sun, J. C. Huffman, D. N. Hendrickson and G. Christou},  Chem. Commun. (Cambridge) 21, 2672 (2003).

\bibitem{foot}
{The noise at low temperatures is exaggerated by two factors: a non-linearity in the thermometer that limits the resolution for temperatures below $0.4K$; and noise associated with digitizing (continuous) data acquired by the oscilloscopes.}

\bibitem{LowTemperaturePhysics}
{C. Enss and S. Hunklinger}, Low-temperature Physics, (Springer-Verlag, Berlin, 2005).

\bibitem{Poiseuille}  Notable exceptions are singular systems in which Poiseuille flow of phonons has been found, for example, in liquid helium in a very narrow temperature range at low temperatures (See Ref. \cite{LowTemperaturePhysics}).

\bibitem{separatespeed}  We did not obtain a reliable determination of the separate avalanche speeds at low magnetic fields.  This is an interesting area for future study.

\bibitem{footII}
{The minor species magnetization, $M_z^{minor}$, also contributes to $B_z$; however, it is neglected since $M^{minor} \ll M^{major}$.}

\end{thebibliography}
\end{document}